\documentclass[useAMS,usenatbib]{mn2e}
\usepackage[dvips]{graphicx}
\title[Type Ibn Supernova SN 2006jc]{Optical photometry and spectroscopy of 
the type Ibn supernova SN 2006jc until the onset of dust formation}

\author[G.C. Anupama et al.]
{G. C. Anupama$^1$, D. K. Sahu$^1$, U. K. Gurugubelli$^{1,2}$, T. P. Prabhu$^1$,\newauthor N. Tominaga$^3$, M. Tanaka$^{4,5}$, K. Nomoto$^{5,4}$\\
1. Indian Institute of Astrophysics, Bangalore, 560 034, India\\
2. Joint Astronomy Programme, Indian Institute of Science, Bangalore 560 012,
India\\
3. Optical and Infrared Astronomy Division, National Astronomical Observatory, 
2-21-1 Osawa, Mitaka, Tokyo 181-8588, Japan\\
4. Department of Astronomy, Graduate School of Science,
University of Tokyo, Hongo 7-3-1, Bunkyo-ku, Tokyo 113-0033, Japan\\
5. Institute for the Physics and Mathematics of the
Universe, University of Tokyo, Kashiwa, Chiba 277-8568, Japan \\
(E-mail:gca{@}iiap.res.in)}

\begin{document}

\date{Accepted.....; Received .....}


\maketitle

\label{firstpage}

\begin{abstract}
We present optical $UBVRI$ photometric and spectroscopic data of the type Ibn
supernova SN 2006jc, until the onset of the dust forming phase. The optical
spectrum shows a blue continuum and is dominated by the presence of moderately
narrow (velocity $\sim 2500$ km s$^{-1}$) He I emission lines superimposed over
a relatively weak supernova spectrum. The helium lines are produced in a
pre-existing He rich circumstellar shell. The observed helium line fluxes 
indicate the circumstellar shell is dense, with a density of 
$\sim 10^9 - 10^{10} $~cm$^{-3}$. The helium mass in this shell is estimated to 
be $\la 0.07$~M$_\odot$. The optical light curves show a clear signature of 
dust formation, indicated by a sharp decrease in the magnitudes around day 50, 
accompanied by a reddening of the colours. The evolution of the optical light 
curves during the early phase and that of the $uvoir$ bolometric light curve 
at all phases is reasonably similar to normal Ib/c supernovae. 

\end{abstract}


\section{Introduction}

Core collapse supernovae (CCSNe) signify the end of the most massive stars. 
They are categorized, spectroscopically, as type II, IIb, Ib and Ic by the 
presence of strong hydrogen, helium and weak hydrogen, helium alone and no 
hydrogen or helium, respectively (see Filippenko 1997 for a review). Type IIn 
are those objects that show narrow hydrogen emission lines as a result of the 
interaction of the supernova ejecta with a dense circumstellar medium (CSM) 
(Schlegel 1990, Chugai \& Danziger 1994). Wolf-Rayet stars, i.e., massive stars that have been 
stripped off their outer hydrogen and/or helium layers due to mass loss during 
the course of their evolution, are believed to be the progenitors of the 
stripped-envelope CCSNe (IIb, Ib, Ic). The possibility of yet another class of 
stripped-envelope CCSNe emerged with the detection of moderately narrow helium
emission lines in SN 1999cq (Matheson et al. 2000), similar to the presence of
narrow hydrogen lines in type IIn. Matheson et al.\ suggested an interaction of
the supernova ejecta with a dense CSM that had little or no hydrogen. SN 2002ao
(Martin et al. 2002, Filippenko \& Chornock 2002) was identified to be
similar to SN 1999cq. The recent discovery of SN 2006jc with strong, 
moderately narrow helium emission lines, and weak hydrogen lines has added
to the list of this new, interesting class of CCSNe, now designated as Ibn
(Pastorello et al. 2008).

Supernova SN 2006jc was discovered by K. Itagaki, at a magnitude of 13.8, on 
Oct 9.75 UT 
on an unfiltered image (Nakano et al. 2006). The non-detection of this object
on an image obtained on Sep 22 suggests the supernova was discovered shortly
after explosion. Based on the presence of strong helium features in
the early spectra, the event was classified to be of type Ib (Fesen et al. 
2006, Crotts et al. 2006). The similarity of SN 2006jc with SNe 1999cq and
2002ao was first noted by Benetti et al. (2006). 

The progenitor of this supernova is believed to have experienced a luminous
outburst, similar to those of luminous blue variables (LBVs) two years prior 
to the supernova event (Nakano et al.\ 2006, Pastorello et al.\ 2007). 
Multiwavelength observations of this supernova have shown it
to be unique in many respects. Early {\it Swift} UVOT observations on 2006 
October 13 by Brown et al. (2006) indicate extremely blue UV-V 
colours. X-ray emission has also been observed by the {\it Swift} (X-Ray 
Telescope) and the {\it Chandra} satellites (Immler et al. 2008). On the 
contrary, SN 2006jc was not detected in the early-time radio observation 
(Soderberg 2006). The X-ray emission and the UV excess are attributed to an 
interaction of the supernova ejecta with a shell of material deposited during 
the recent luminous outburst of the progenitor (Immler et al. 2008).

The optical spectra (Pastorello et al. 2007, Foley et al. 2007) show a 
prominent blue continuum that lasts well into the onset of the nebular phase. 
Around 75 days since maximum, the 
steepness of the blue continuum had dropped, while the continuum in the red 
had brightened, with the overall spectrum taking a  "U"-shape (Smith et al. 
2008). The red excess had disappeared by day 128. Interestingly, the optical
light curves showed a sharp decline after day 50, while the near infrared
luminosities brightened during the same epoch (Arkharov et al. 2006,  Minezaki
et al. 2007; Smith et al. 2008). The NIR excess peaked around 80 days and
persisted to past 200 days (Di Carlo et al. 2008, Mattila et al. 2008; Sakon
et al. 2007). Smith et al., Di Carlo et al, and Mattila et al., attribute this
NIR excess to the formation of a hot dust in an outward shock-formed cool dense
shell, while Sakon et al. (2008), Nozawa et al. (2008) and Tominaga et al. 
(2008) propose dust formation in the supernova ejecta.

We present in this paper the optical photometric and spectroscopic observations 
of SN 2006jc obtained with the 2m Himalayan Chandra Telescope during 2006 
October 14 -- 2007 January 13. 

\section{Observations}
 
\begin{figure}
\centering
\includegraphics[width=\columnwidth]{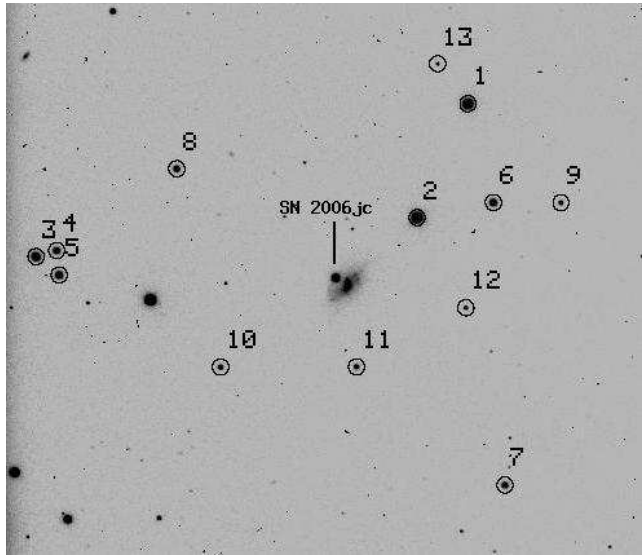}
\caption{The field of SN 2006jc. Stars used as secondary standards and listed
in Table 1 are marked.}
\label{field}
\end{figure}

Photometric monitoring of SN 2006jc in the $U B V R$ and $I$ bands began on
2006 October 16 (JD 2454025.45) and continued until 2007 January 13 
(JD 2454114.40), using the Himalaya Faint Object Spectrograph Camera (HFOSC).
Photometric standard fields (Landolt 1992) PG0231+051 and PG0942-029 were observed 
on 2006 November 23 and the fields PG1047+003, PG0942-029 and PG1323-086 were observed 
on 2006 December 27 under photometric conditions. These were used to calibrate a 
sequence of secondary standards in the supernova field. The data reduction and 
photometry was done in the standard manner, using the various tasks available 
within IRAF. The observed data were bias subtracted,
flat-field corrected and cosmic-ray hits removed. Aperture photometry was
performed on the standard stars, using an aperture radius determined using the
aperture growth curve, and were calibrated using the average colour terms and
photometric zero points determined on the individual nights. The $UBVRI$
magnitudes of the secondary standards in the supernova field, calibrated and
averaged over the two nights are listed in Table \ref{stdmag}. The secondary 
sequence is shown in Figure \ref{field}, marked by numbers. The magnitudes of 
the supernova and the local standards were estimated using the profile fitting 
technique, using a fitting radius equal to the FWHM of the stellar profile. The 
difference between the aperture and profile-fitting magnitude (aperture 
correction) was obtained 
using the bright standards in the supernova field and this correction was 
applied to the supernova magnitude. The calibration of the supernova magnitude 
to the standard system was done differentially with respect to the local 
standards.

\begin{table*}
\caption{Magnitudes for the sequence of secondary standard stars in the field
of SN 2006jc. The stars are identified in Fig. \ref{field}.}
\centering
\begin{tabular}{lccccc}
\hline\hline
ID & U  & B & V &  R & I \\
\hline
1  &     $13.636\pm0.020$& $13.482\pm0.005$& $12.884\pm0.010$&
$12.537\pm0.031$&   $12.156\pm0.015$\\
2  &     $13.005\pm0.020$& $13.167\pm0.005$& $12.795\pm0.011$&
$12.528\pm0.020$&   $12.202\pm0.020$\\
3  &     $14.663\pm0.030$& $14.280\pm0.005$& $13.534\pm0.013$&
$13.175\pm0.029$&   $12.699\pm0.053$\\
4  &     $15.511\pm0.050$& $15.376\pm0.005$& $14.749\pm0.011$&
$14.397\pm0.006$&   $14.030\pm0.050$\\
5  &     $15.084\pm0.030$& $14.966\pm0.005$& $14.307\pm0.010$&
$13.937\pm0.008$&   $13.543\pm0.004$\\
6  &     $15.142\pm0.030$& $15.100\pm0.011$& $14.503\pm0.005$&
$14.102\pm0.003$&   $13.722\pm0.010$\\
7  &     $16.140\pm0.037$& $15.742\pm0.012$& $14.994\pm0.005$&
$14.553\pm0.005$&   $14.145\pm0.005$\\
8  &     $16.068\pm0.033$& $15.614\pm0.005$& $14.739\pm0.008$&
$14.246\pm0.003$&   $13.754\pm0.013$\\
9  &     $17.974\pm0.041$& $17.163\pm0.020$& $16.240\pm0.006$&
$15.712\pm0.005$&   $15.277\pm0.014$\\
10 &     $16.862\pm0.046$& $16.695\pm0.010$& $16.055\pm0.005$&
$15.661\pm0.005$&   $15.255\pm0.006$\\
11 &     $16.590\pm0.032$& $16.705\pm0.016$& $16.207\pm0.005$&
$15.860\pm0.005$&   $15.519\pm0.015$\\
12 &     $18.332\pm0.058$& $17.462\pm0.013$& $16.545\pm0.016$&
$16.005\pm0.007$&   $15.549\pm0.031$\\
13 &     $19.724\pm0.070$& $18.189\pm0.008$& $16.912\pm0.023$&
$16.127\pm0.007$&   $15.472\pm0.028$\\
\hline
\multicolumn{6}{l}{The errors quoted are the statistical errors associated
with the magnitudes.}\\
\end{tabular}
\label{stdmag}
\end{table*}

Spectroscopic monitoring of SN 2006jc began on 2006 October 14 (JD 2454023.47)
and continued until 2006 Dec 14 (JD 2454084.35). The log of spectroscopic 
observations is given in Table \ref{speclog}. The data reduction was
carried out in the standard manner using the tasks available within IRAF. The
data were bias corrected, flat-fielded and the one dimensional spectra
extracted using the optimal extraction method. Spectra of FeAr and FeNe lamps
were used for wavelength calibration. The instrumental response curves were
obtained using spectrophotometric standards observed on the same night and the
supernova spectra were brought to a relative flux scale. The flux calibrated
spectra in the two regions were combined to a weighted mean to give the final
spectrum on a relative flux scale. The spectra were brought to an absolute flux
scale using zero points derived by comparing the observed flux in the spectra
with the flux estimated using the photometric magnitudes.

\begin{table}
\caption{Log of spectroscopic observations of SN 2006jc}
\begin{tabular}{lcrcc}
\hline\hline
Date & J.D. & \multicolumn{1}{c}{Phase{\rlap{*}}} & Resln & Range \\
     & 2450000+ & \multicolumn{1}{c}{(days)} & \AA\ & \AA\ \\
\hline
2006 Oct 14 & 4023.47 &  7.5    & 7 & 3500-7800; 5200-9250\\
2006 Oct 16 & 4025.49 &  9.5    & 7 & 3500-7800; 5200-9250\\
2006 Oct 17 & 4026.44 &  10.5    & 7 & 3500-7800; 5200-9250\\
2006 Oct 20 & 4029.49 &  13.5    & 7 & 3500-7800; 5200-9250\\
2006 Oct 23 & 4032.46 &  16.5   & 7 & 3500-7800; 5200-9250\\
2006 Oct 24 & 4033.43 &  17.5   & 7 & 3500-7800; 5200-9250\\
2006 Oct 29 & 4038.45 &  22.5   & 7 & 3500-7800; 5200-9250\\
2006 Oct 30 & 4039.48 &  23.5   & 7 & 3500-7800; 5200-9250\\
2006 Nov 01 & 4041.48 &  25.5   & 7 & 3500-7800; 5200-9250\\
2006 Nov 05 & 4045.37 &  29.4   & 7 & 3500-7800; 5200-9250\\
2006 Nov 07 & 4047.42 &  31.4   & 7 & 3500-7800; 5200-9250\\
2006 Nov 14 & 4054.45 &  37.5   & 9 & 3500-7800\\
2006 Nov 16 & 4055.51 &  39.5   & 7 & 3500-7800; 5200-9250\\
2006 Nov 18 & 4058.41 &  42.4   & 7 & 3500-7800; 5200-9250\\
2006 Nov 23 & 4063.34 &  47.3   & 7 & 3500-7800; 5200-9250\\
2006 Dec 14 & 4084.35 &  68.4   & 9 & 3500-7800\\
\hline
\multicolumn{5}{l}{{\rlap{*}}\ \ \ With respect to date of maximum, assumed to be}\\
\multicolumn{5}{l}{\, JD\,2454016 (Pastorello et al. 2007)}
\end{tabular}
\label{speclog}
\end{table}

\section{Results}

\subsection{Light curves}

\begin{figure}
\centering
\includegraphics[width=\columnwidth]{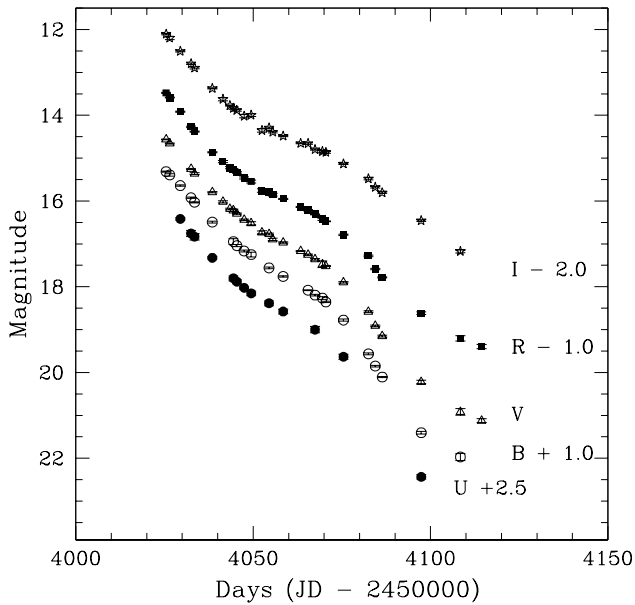}
\caption{$UBVRI$ light curves of SN 2006jc obtained with the HCT.}
\label{phot1}
\end{figure}

\begin{table*}
\caption{Photometric observations of SN 2006jc}
\begin{tabular}{lccccccc}
\hline\hline
Date & J.D. & Phase\rlap{*} & U &  B & V &  R &  I\\
     & 2450000+ & (days) & \\
\hline
2006 Oct 16 &  4025.45& 9.45 & $	           $&  $14.314\pm0.014$&	$14.563\pm 0.011$&	$14.480\pm0.013$&$14.107\pm 0.015$\\
2006 Oct 17 &  4026.49& 10.49 & $	           $&  $14.392\pm0.026$&	$14.659\pm 0.013$&	$14.587\pm0.010$&$14.200\pm0.037$\\
2006 Oct 20 &  4029.46& 13.46 & $13.918\pm0.027$&  $14.641\pm0.015$&	$	        $&	$14.918\pm0.013$&$14.508\pm0.015$\\
2006 Oct 23 &  4032.48& 16.48& $14.256\pm0.070$&  $14.925\pm0.016$&	$15.252\pm 0.011$&	$15.271\pm0.017$&$14.799\pm0.018$\\
2006 Oct 24 &  4033.48& 17.48& $14.333\pm0.074$&  $15.030\pm0.012$&	$15.360\pm 0.023$&	$15.379\pm0.012$&$14.897\pm0.022$\\
2006 Oct 29 &  4038.49& 22.49& $14.824\pm0.025$&  $15.493\pm0.020$&	$15.794\pm 0.013$&	$15.864\pm0.012$&$15.373\pm0.015$\\
2006 Nov 01 &  4041.49& 25.49& $	           $&  $	      $&	$16.018\pm 0.023$&	$16.079\pm0.026$&$15.626\pm0.036$\\
2006 Nov 03 &  4043.46& 27.46& $	           $&  $	      $&	$16.187\pm 0.017$&	$16.220\pm0.026$&$15.774\pm0.014$\\
2006 Nov 04 &  4044.44& 28.44& $15.304\pm0.056$&  $15.947\pm0.069$&	$16.218\pm 0.026$&	$16.265\pm0.025$&$15.836\pm0.026$\\
2006 Nov 05 &  4045.41& 29.41& $15.381\pm0.047$&  $16.041\pm0.025$&        $16.294\pm 0.017$&      $16.334\pm0.019$&$15.883\pm0.021$\\
2006 Nov 07 &  4047.45& 31.45& $15.524\pm0.028$&  $16.167\pm0.022$&	$16.438\pm 0.017$&	$16.465\pm0.025$&$16.019\pm0.038$\\
2006 Nov 09 &  4049.44& 33.44& $15.654\pm0.051$&  $16.249\pm0.055$&	$16.520\pm 0.035$&	$16.545\pm0.033$&$15.996\pm0.030$\\
2006 Nov 12 &  4052.49& 36.49& $	           $&  $	      $&	$16.732\pm 0.036$&	$16.756\pm0.043$&$16.353\pm0.040$\\
2006 Nov 14 &  4054.48& 38.48& $15.885\pm0.055$&  $16.559\pm0.021$&	$16.771\pm 0.020$&	$16.794\pm0.017$&$16.295\pm0.017$\\
2006 Nov 15 &  4055.52& 39.52& $	           $&  $	      $&	$16.894\pm 0.025$&	$16.853\pm0.029$&$16.392\pm0.020$\\
2006 Nov 18 &  4058.46& 42.46& $16.078\pm0.053$&  $16.760\pm0.017$&	$16.969\pm 0.015$&	$16.950\pm0.011$&$16.484\pm0.013$\\
2006 Nov 23 &  4063.38& 47.38& $	           $&  $	      $&	$17.168\pm 0.010$&	$17.140\pm0.010$&$16.655\pm0.013$\\
2006 Nov 25 &  4065.47& 49.47& $	           $&  $17.079\pm0.010$&	$17.254\pm 0.016$&	$17.214\pm0.014$&$16.658\pm0.017$\\
2006 Nov 27 &  4067.44& 51.44& $18.472\pm0.074$&  $17.195\pm0.019$&	$17.360\pm 0.017$&	$17.312\pm0.011$&$16.800\pm0.015$\\
2006 Nov 29 &  4069.51& 53.51& $	           $&  $17.263\pm0.029$&	$17.474\pm 0.078$&	$17.429\pm0.025$&$16.844\pm0.031$\\
2006 Nov 30 &  4070.50& 54.50& $              $&  $17.355\pm0.015$&	$17.517\pm 0.012$&	$17.473\pm0.011$&$16.865\pm0.024$\\
2006 Dec 05 &  4075.45& 59.45& $17.132\pm0.061$&  $17.777\pm0.024$&	$17.898\pm 0.016$&	$17.788\pm0.013$&$17.133\pm0.017$\\
2006 Dec 12 &  4082.49& 66.49& $	           $&  $18.565\pm0.021$&	$18.580\pm 0.020$&	$18.281\pm0.019$&$17.482\pm0.026$\\
2006 Dec 14 &  4084.42& 68.42& $	           $&  $18.850\pm0.019$&	$18.912\pm 0.021$&	$18.587\pm0.014$&$17.680\pm0.017$\\
2006 Dec 16 &  4086.39& 70.39& $	           $&  $19.106\pm0.014$&        $19.150\pm 0.020$&      $18.780\pm0.014$&$17.805\pm0.019$\\
2006 Dec 27 &  4097.38& 81.38& $19.935\pm0.052$&  $20.404\pm0.033$&        $20.220\pm 0.039$&      $19.617\pm0.030$&$18.458\pm0.021$\\
2007 Jan 07 &  4108.42& 92.42& $	           $&  $20.971\pm0.082$&	$20.920\pm 0.077$&	$20.202\pm0.064$&$19.173\pm0.036$\\
2007 Jan 13 &  4114.40& 98.40& $              $&  $              $&        $21.122\pm 0.046$&      $20.394\pm0.050$&$              $\\
\hline
\multicolumn{8}{l}{\rlap{*}\ \ \ With respect to date of maximum, assumed to
be JD\,2454016.}\\
\multicolumn{8}{l}{The errors quoted are the statistical errors associated
with the magnitudes.}\\
\end{tabular}
\label{phot}
\end{table*}

\begin{figure}
\centering
\includegraphics[width=\columnwidth]{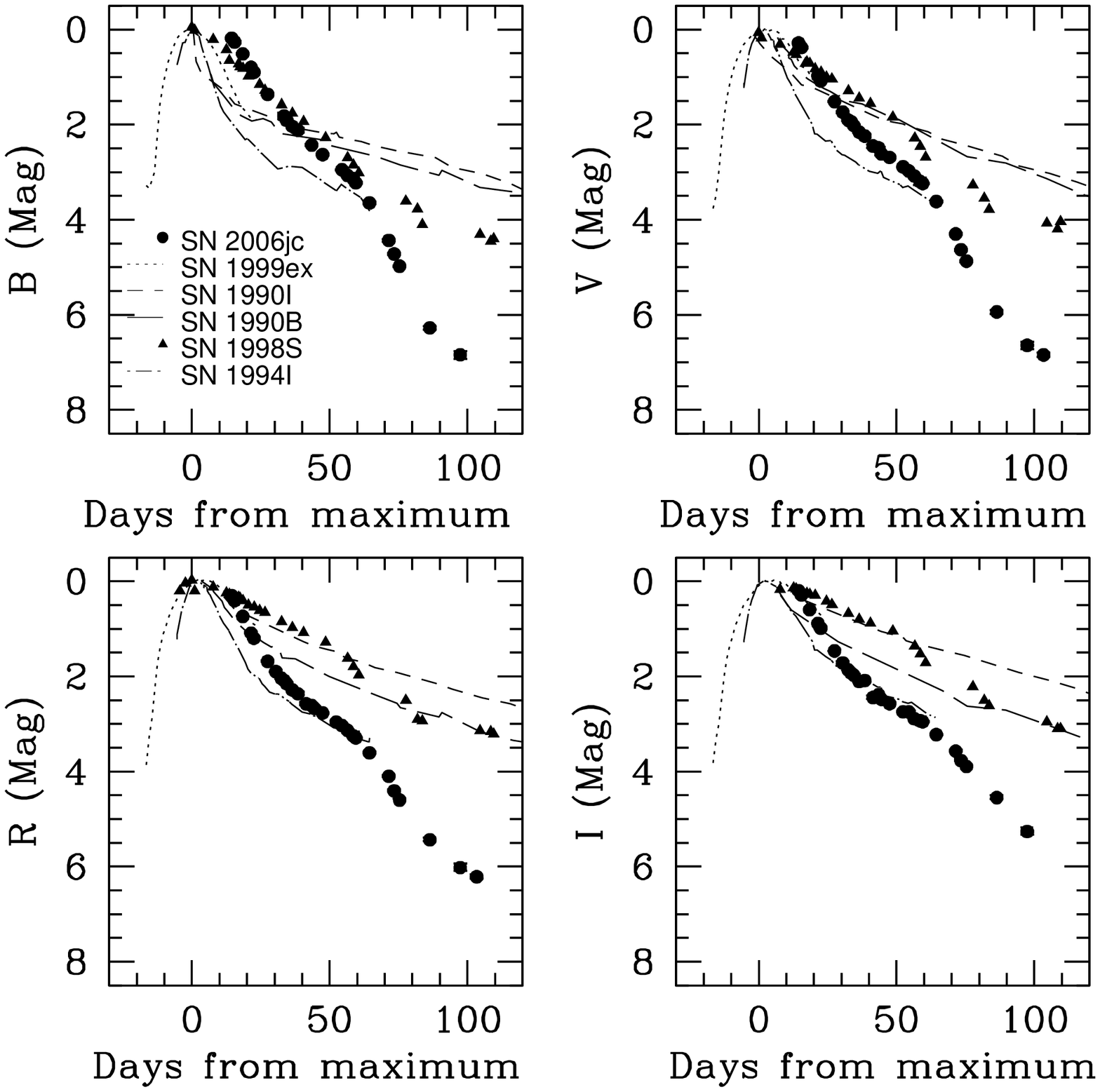}
\caption{Comparison of the $BVRI$ light curves with SNe 1999ex (Ib/c), SN 1994I
(Ic), SN 1990I (Ib), SN 1990B (Ic), and SN 1998S (IIn). The magnitudes of 
SN 2006jc are normalized with respect to those on JD 454021.7 ($B=14.13$, 
$V=14.28$, $R=14.18$ and $I=13.91$: Pastorello et al.\ 2007), while the 
magnitudes of other SNe are normalized with respect to the magnitudes at their 
respective maximum (see text).}

\label{phot2}
\end{figure}

The $UBVRI$ magnitudes of SN 2006jc are tabulated in Table \ref{phot} and 
plotted in Figure \ref{phot1}. Figure \ref{phot2} shows the light curves of 
SN 2006jc compared with
the light curves of a few SNe Ib/c, namely, SN 1999ex (Ib/c) (JD$_{\rm{max,B}}
=245\,1498.1$: Stritzinger et al.\ 2002),
SN 1994I (Ic) (JD$_{\rm{max,B}}=244\,9450.56$: Richmond et al. 1996), SN 1990I 
(Ib) (JD$_{\rm{max}}=244\,8010$: Elmhamdi et al. 2004) and 
SN 1990B (Ic) (JD$_{\rm{max,B}}=244\,7909.0$: Clocchiatti et al. 2001), and the 
type IIn SN 1998S (JD$_{\rm{max,R}}=245\,0890.0$: Fassia et al. 2000). The 
early 
decline ($\sim 10-25$~days; JD(max)=2454016{\footnote{Comparing the early spectra of
SN 2006jc with those of the type Ibn SN 2000er, Pastorello et al. (2008) suggest that 
SN 2006jc was discovered $\sim 10$ days after maximum light, estimated to have
occured on JD $2450008\pm 15$. Since the uncertainity in the estimated epoch of
maximum is quite large, we choose to follow the epoch estimated by Pastorello
et al. (2007). }}: Pastorello et al. 2007) in the $U$ and 
$B$ bands is similar to SN 1999ex,
while the decline is very similar to SN 1994I in the $VRI$ bands. The R 
band light curve of SN 2006jc shows a decline during days $\sim 10-20$ which 
is slower compared with that of the type Ibn SN 1999cq (Matheson et al. 2000).
Between days 30--50, a flattening is seen in all the bands, and the decline rate
appears very similar to SN 1990B and SN 1990I. However, beyond day 50, SN 2006jc
behaves in a quite different manner compared to the other SNe. A sharp decline 
is seen in all the bands, with the decline being the steepest in the $U$ and 
$B$ bands. 

\begin{figure}
\centering
\includegraphics[width=\columnwidth]{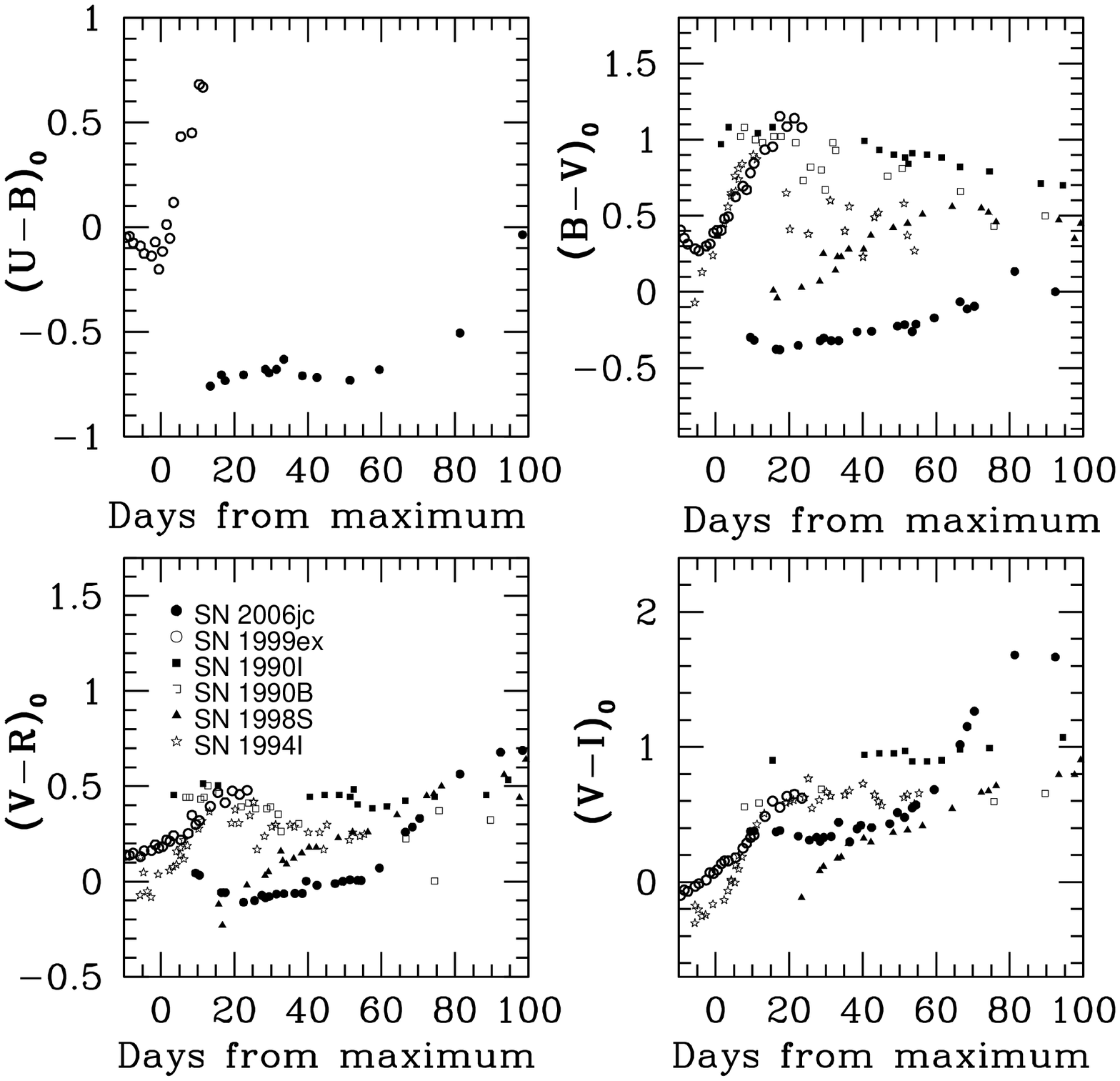}
\caption{Reddening corrected $U-B$, $B-V$, $V-R$ and $R-I$ colour curves of
SN 2006jc compared with other type Ib, Ic and IIn supernovae.}
\label{phot3}
\end{figure}

The $U-B$, $B-V$, $V-R$ and $V-I$ colours of SN 2006jc, reddening
corrected using $E(B-V)=0.05$ (Pastorello et al. 2007) are shown in Figure 
\ref{phot3}. Also shown in the Figure are the colours for SN 1999ex, SN 1994I, 
SN 1990I, SN 1990B and SN 1998S reddening corrected using $E(B-V)$ values 0.30
(Stritzinger et al. 2002), 0.45 (Richmond et al. 1996), 0.13 (Elmhamdi et al.
2004), 0.85 (Clocchiatti et al. 2001) and 0.22 (Fassia et al. 2000), 
respectively. The $U-B$, $B-V$, $V-R$ and $V-I$ 
colour curves show very little colour evolution until $\sim$ day 60. SN 2006jc 
had significantly bluer $U-B$, $B-V$ and $V-R$ colours compared to other 
supernovae, while the $V-I$ colour is similar to other SNe. A dramatic change 
in all the colours is seen 
beyond day 60. SN 2006jc begins to get redder, by $\sim 0.5$ mag. Maximum change
is seen in $V-I$ colour, which changes by almost 1 mag by day 100. The 
reddening of colours observed in SN 2006jc at later phases is not seen in 
other SNe, except in the case of the IIn SN 1998S.

\begin{figure}
\centering
\includegraphics[width=\columnwidth]{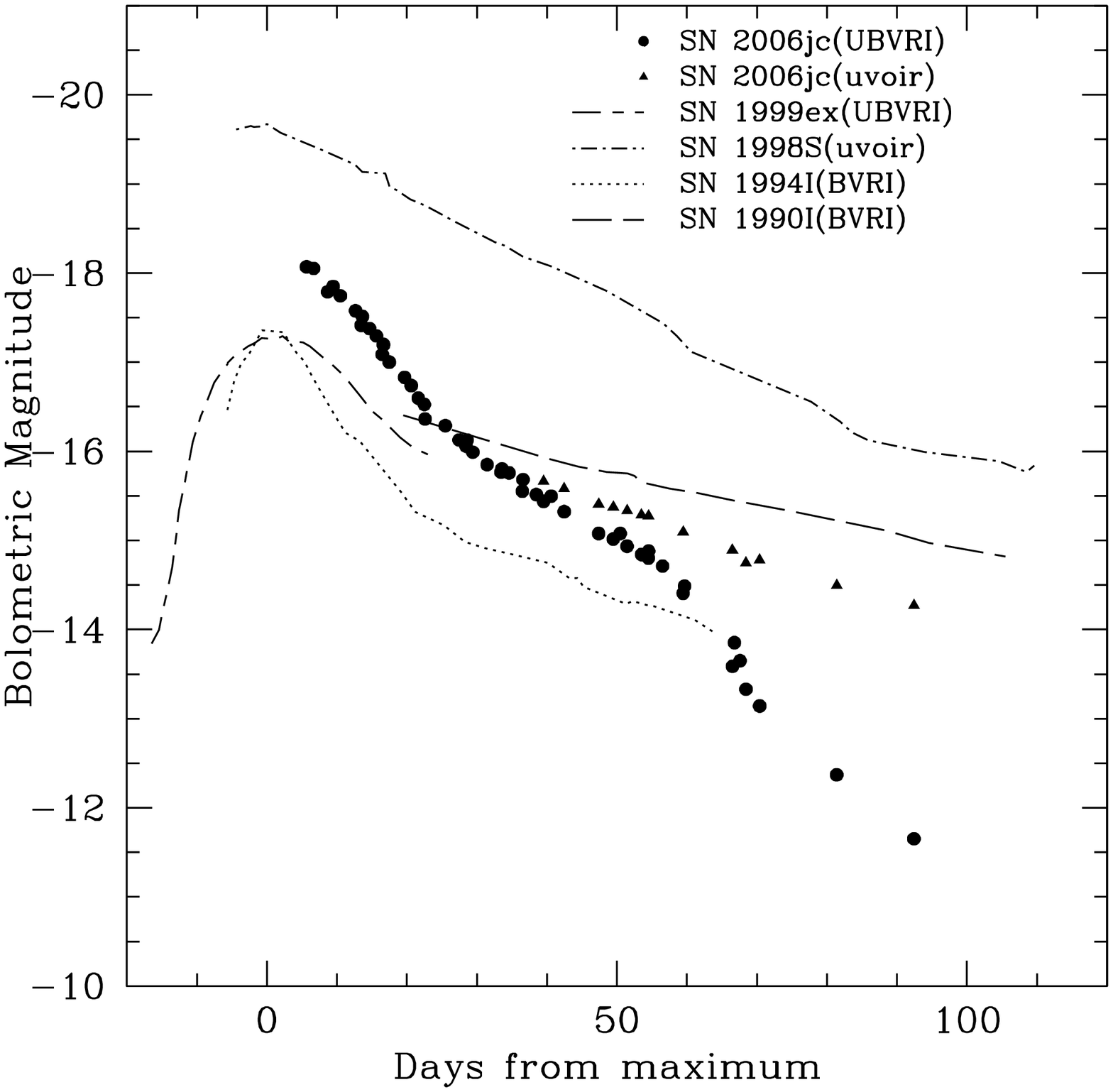}
\caption{The optical `quasi-bolometric' and the $uvoir$ bolometric light curves
of SN 2006jc. Also shown in the figure are the $UBVRI$ bolometric light curves 
of the type Ib/c SN 1999ex and SN 1994I and the $uvoir$ bolometric light
curve of the type IIn SN 1998S.}
\label{bol}
\end{figure}

The sudden decline in the UBVRI light curves, coincident with the increase in 
the luminosities in the near-IR region (Arkharov et al. 2006), with a reddening
of the colours are quite similar to the behaviour of dust forming novae 
(e.g.\ Gehrz 1988) and a clear indication of the formation of hot dust. 

The optical `quasi-bolometric' light curve is constructed using the $UBVRI$
magnitudes presented here, and those published by Pastorello et al.\ (2007).
Assuming a distance of 25.8 Mpc and a reddening of $E(B-V) = 0.05$ (Pastorello
et al.\ 2007), the observed UBVRI magnitudes were converted to monochromatic
fluxes and integrated over the observed wavelength range to obtain the optical
quasi-bolometric light curve. Likewise, combining the NIR magnitudes (Arkharov
et al.\ 2006, Di Carlo et al.\ 2007, Mattila et al.\ 2008) with the UBVRI
magnitudes, the optical+NIR ($uvoir$) bolometric light curve was constructed 
integrating over the $U$ to $K$ bands (see also Pastorello et al. 2007, 
Mattila et al. 2008, Di Carlo et al. 2008, Tominaga et al. 2008). Figure \ref{bol} shows
the optical `quasi-bolometric' light curve as well as the $uvoir$ bolometric
light curve. Also shown in the Figure are the bolometric light curves of type
Ib/c SNe 1990I, 1994I and 1999ex and the type IIn SN 1998S. It is evident from 
the plot that SN 2006jc has a luminosity that is higher than that of other 
Ib/c SNe, while it is about 1.6 magnitudes fainter than the type IIn SN 1998S.
The optical luminosity
of SN 2006jc indicates an early decline rate of 0.092 mag day$^{-1}$ until 
$\sim 30$ days since maximum. A flattening is seen in the light curve during 
$\sim 30-50$ days after maximum, with the decline rate during this period 
being 0.047 mag day$^{-1}$. The onset of dust formation is marked by a sharp 
decline in the optical luminosity, with a decline rate of 0.084 mag day$^{-1}$ 
during $\sim 50-100$ days after maximum. In contrast to the optical luminosity, 
the $uvoir$ bolometric luminosity shows a flat decline, with a decline rate of 
0.026 mag day$^{-1}$ beyond day 35. A comparison with the bolometric light 
curves of other SNe indicates that the bolometric light curve decline of 
SN 2006jc is not too different from other normal SNe Ib/c. The early decline 
lies between the rapidly declining SN 1994I (0.106 mag day$^{-1}$) 
and the slower SN 1999ex (0.076 mag day$^{-1}$), while the $uvoir$ decline 
rate at later phases is similar to SN 1994I (0.029 mag day$^{-1}$).

\subsection{The spectrum and its evolution}

\begin{figure}
\centering
\includegraphics[width=\columnwidth]{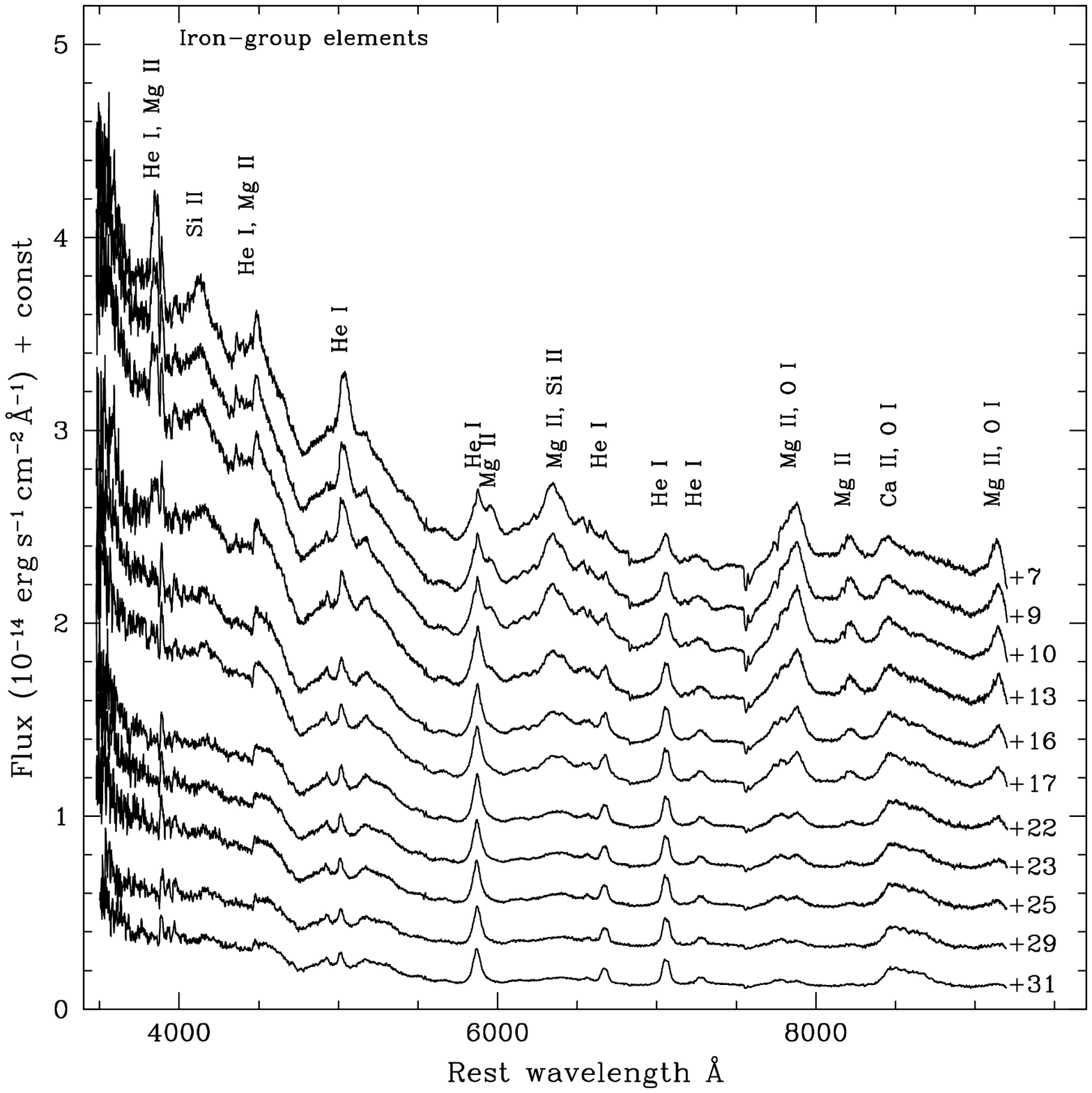}
\caption{Spectroscopic evolution of SN 2006jc during +7--+31 days since 
maximum on JD 2454016. Note the fading of the broad features, the evolution of
Fe II lines in the 4500--5500\AA\ region, and the development of H$\alpha$ line
from an absorption feature into an emission feature.}
\label{spec1}
\end{figure}

\begin{figure}
\centering
\includegraphics[width=\columnwidth]{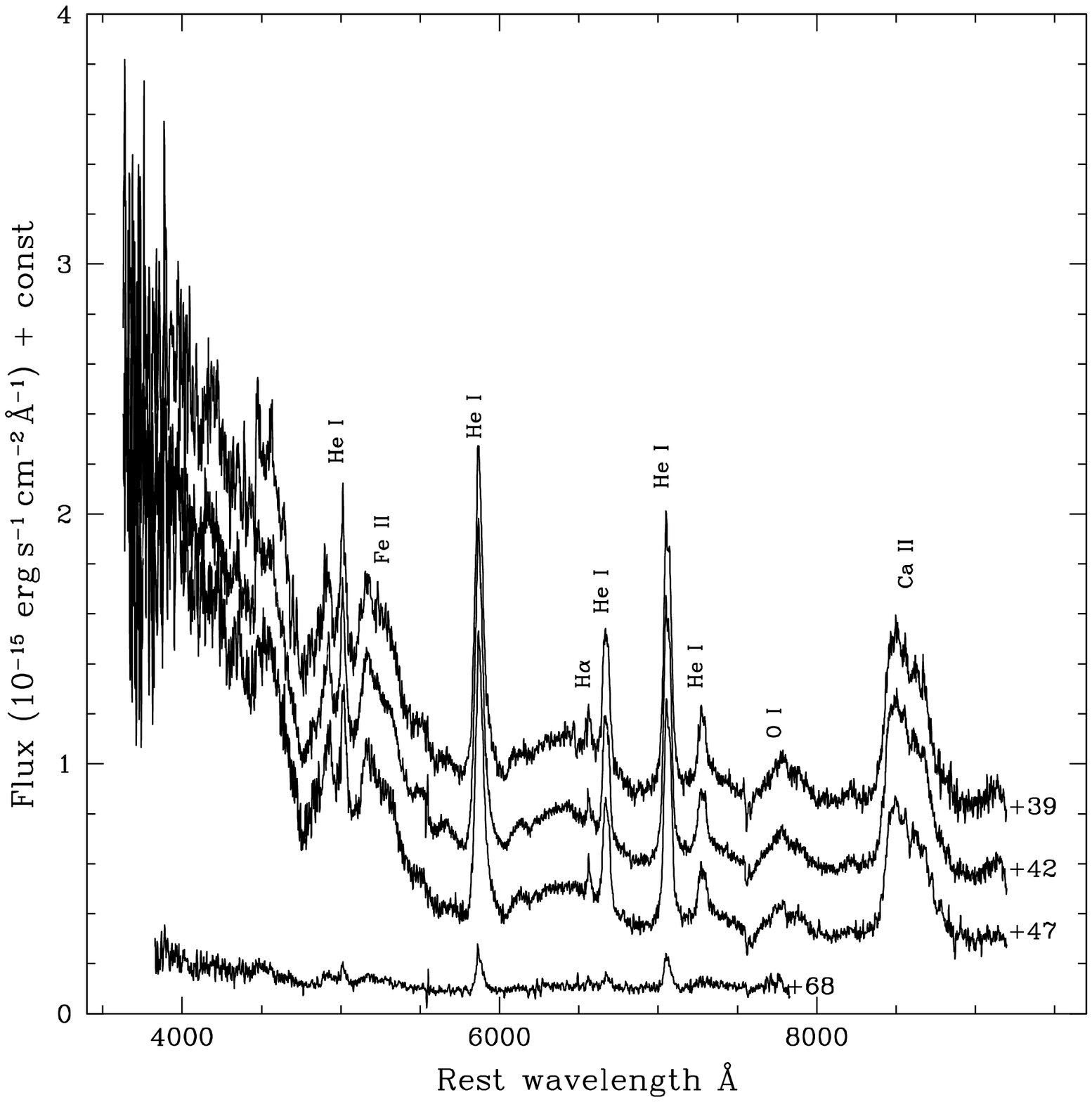}
\caption{Spectroscopic evolution of SN 2006jc during +39--+68 days since 
maximum on JD 2454016. Note the increase in strength of the O I 7774 \AA\ and
Ca II infrared triplet lines.}
\label{spec2}
\end{figure}

The spectrum of SN 2006jc and its evolution during the phase +7 to +68 days 
since the estimated maximum on JD 2454016 is presented in Figures \ref{spec1} 
and \ref{spec2}. These spectra provide a fairly dense coverage of the 
early-time spectral evolution of SN 2006jc and are complementary in phase to 
those presented by Foley et al. (2007), Smith et al. (2008) and Pastorello et 
al. (2007, 2008). The 
spectrum is peculiar and different from that of normal type Ib/c supernovae 
(Matheson et al. 2001, Branch et al. 2002). The photospheric P Cygni profiles
that are typically found in the early spectra of type Ib/c supernovae are
absent, and the spectrum is characterized by (a) a steep, blue continuum 
shortward of $\sim 5500$~\AA\ and (b) dominant moderately narrow helium 
emission lines. 

\begin{figure}
\centering
\includegraphics[width=\columnwidth]{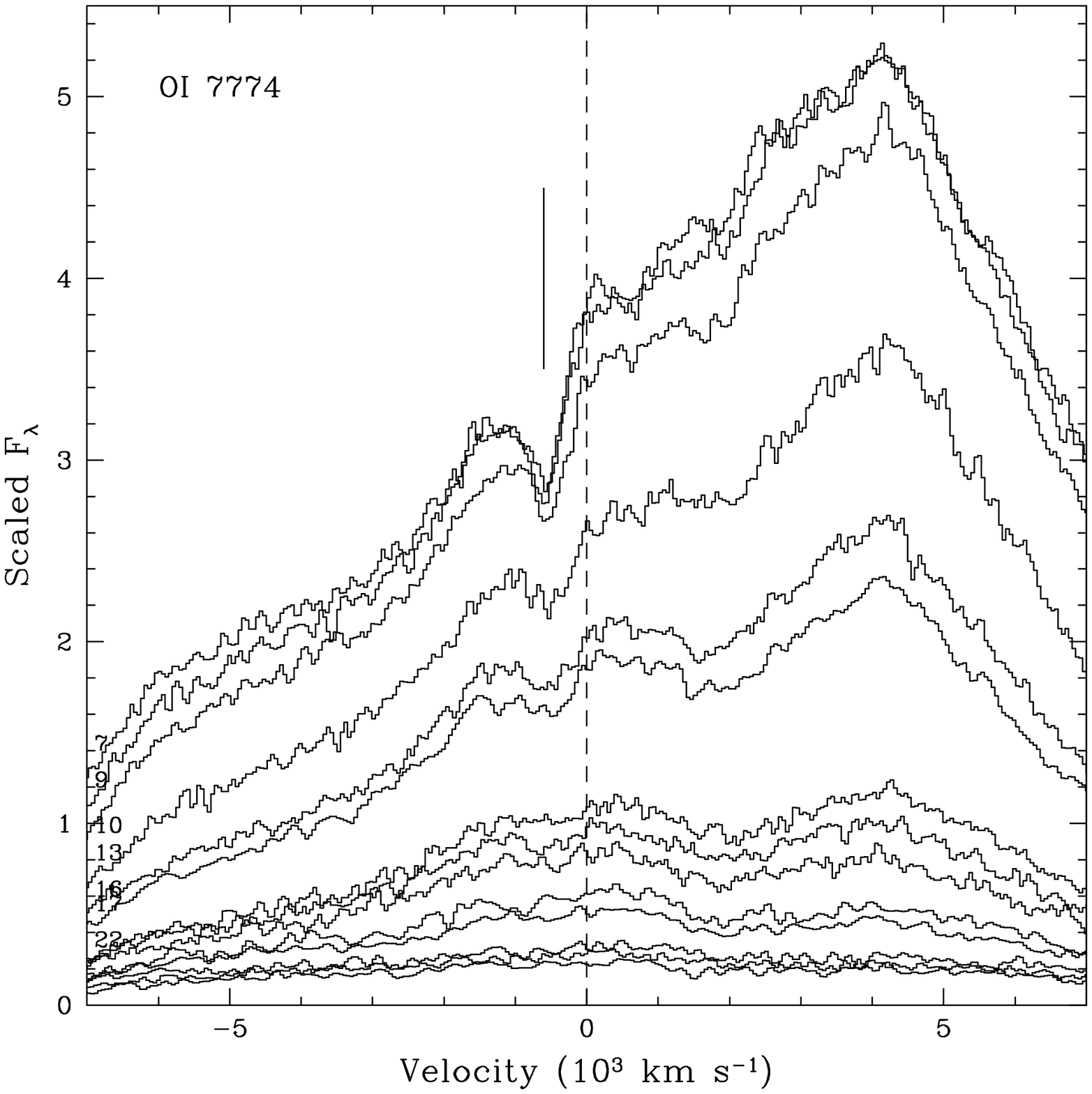}
\caption{The OI 7774 \AA\ line profile. Note the sharp absorption at $\sim 620$
km s$^{-1}$ (marked by the short vertical line) on days 7, 9 and 10, and its
subsequent fading.}
\label{oi}
\end{figure}

Broad emission features, due to the expanding supernova 
material are also seen at $\sim$ 3850, 4121, 5950, 6348, 7800, 8214 and 8500 
\AA. The features at $\lambda\lambda$ 3850, 5950, 7800 and 8214 are identified
with Mg II 3848, 3850 \AA, 5938, 5943 \AA, 7790, 7877 \AA, and 8214, 8234 \AA\
respectively. The feature at 6348 \AA\ is identified with Si II $\lambda$ 6355,
probably blended with Mg II 6346 \AA, and the feature are 4121 \AA\ could be
due to Si II 4128, 4131 \AA. Si II $\lambda\lambda$ 5041, 5056 could
also be present. O I 7774 \AA\ is also present, blended with Mg II. The feature
at 8500 \AA\ is due to Ca II infrared triplet. Mg II 4481 \AA\ could be blended 
with He I 4471 \AA. The full width at half maximum of the broad features 
indicate a velocity of $\sim 5000-6000$ km s$^{-1}$. 
The strength of the Mg II and Si II features decrease with time, while that of
the Ca II IR triplet and O I 7774 \AA\ increase with time. Fe II features 
appear to develop in the 4500-5500 \AA\ region, around 10 days after B 
maximum. The O I 7774 \AA\ line shows a sharp P Cygni absorption during the
early phases, with the absorption minimum indicating a velocity of $\sim 620$
km s$^{-1}$ (Figure \ref{oi}).

The He I line widths indicate a velocity of $\sim 2200-3000$ km s$^{-1}$. As
noted by Foley et al. (2007), a narrow P Cygni component is noticed in the
He I 3889 \AA\ line, at a velocity of $\sim 620$ km s$^{-1}$. A similar component
could be present in the 4471 \AA\ line also, at $\sim 670$ km s$^{-1}$. 
It is interesting to note that the velocity of the narrow P Cygni 
absorption seen in the O I 7774 \AA\ line is very similar to the velocity of
this component.

\begin{figure}
\centering
\includegraphics[width=\columnwidth]{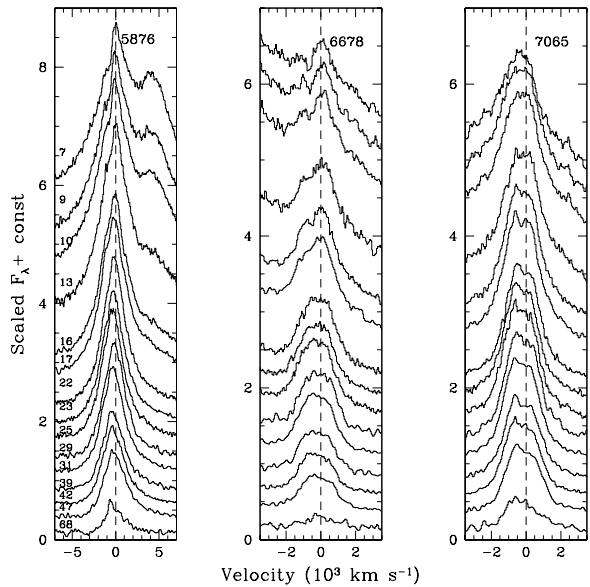}
\caption{Evolution of the line profiles of the He I 5876, 6678 and 7065 \AA\
lines.}
\label{heprof}
\end{figure}

The line
profiles of the strongest He I lines at 5876, 6678 and 7065 \AA\ are shown in
Figure \ref{heprof}. From the figure, it appears that He I 5876 \AA\ may have
a contribution from the fast moving supernova material. A broad component 
could be present in the 5876\AA\ profile in the spectra of days 3, 5, 6 and 
9. A simple de-blending of the components, assuming a simple Gaussian profile 
for both components, indicates the broad component has a velocity $\sim 5000$ 
km s$^{-1}$, similar to the supernova features. The broad component fades with 
time, in accordance with the evolution of the Mg II and Si II features from 
the supernova material. Pastorello et al (2008) note the presence of helium in 
the fast moving supernova ejecta in the early phase spectra of SN 2000er which 
has been found to be very similar to SN 2006jc. This broad feature could
also be due to Na I 5893 \AA\ (Pastorello et al. 2007).
The peak of the moderately
narrow 5876 \AA\ component shifts blueward around day 10, and the line profile 
begins to show asymmetry that is clearly seen by day 20. A similar shift in the 
peak and asymmetry is seen in the 6678\AA\ line also, while the peak is 
blueshifted in the 7065\AA\ line all through. The 7065\AA\ line shows a clear
double-peaked structure after day 12. The asymmetry in the line profiles at
later phases was first noted by Smith et al. (2008) also, who find the
asymmetry more pronounced in the spectra obtained at phases later than presented
here, with the strongest decrease in the red side of the lines occuring between
$\sim 50-75$ days, at the same time during which the red/IR continuum appeared
and increased giving rise to a "U-shaped" continuum.

H$\alpha$ is clearly detected in all the spectra presented here, and is seen to
increase in strength at the later phases. While it is
seen in absorption during the early days, it evolves into an emission feature
around day 18--20 (refer Figs. \ref{spec1},\ref{spec2}). Pastorello et al
(2008) note that the supernova features faded completely beyond $\sim 100$ days
and the spectrum was completely dominated by narrow He I circumstellar lines. 
They also note that the strength of H$\alpha$ is almost comparable to that of
He I 6678\AA\ line at this phase. 

\subsection{The Helium emitting region}

The progenitor of SN 2006jc was observed to undergo a luminous mass loss
episode, similar to the giant eruptions seen in LBVs, lasting about 10 days, 
two years prior to the supernova event (Nakano et al.\ 2006, Pastorello et al.\ 
2007). If we assume that the helium emission lines seen in the supernova spectra
arise in the shell ejected during the luminous event, and this shell is helium 
enriched, then, the observed helium line luminosity may be used to estimate the 
density and mass of the shell. The observed, reddening corrected He I 5876 \AA\ 
and 7065 \AA\ line fluxes are listed in Table \ref{heflx}.

If we assume the circumstellar shell (CS) ejected during the luminous mass loss
episode had a velocity of 600 km s$^{-1}$, similar to the velocity of the 
narrow
P Cygni absorption seen in the He I and O I lines, the radius of the shell 
would be $3.8\times 10^{15}$~cm. The corresponding radius for a CS velocity of 
2500 km s$^{-1}$, as observed for the He I emission lines is $1.6\times 10^{16}$~cm.
We also assume the thickness of the shell to be constant and determined by the 
observed duration ($\sim 10$ days) of the luminous mass loss episode (Nakano et 
al.\ 2006).
Using the observed reddening corrected line fluxes and a distance of 25.8 Mpc, 
we estimate an average density in the range $(0.54-3.7)\times 10^{10}$ cm$^{-3}$
for the shell with a velocity of 600 km s$^{-1}$. The corresponding mass range 
for this shell is $M_{\rm{He}} = 0.001-0.008$ M$_\odot$. Likewise, for a 
shell velocity of 2500 km s$^{-1}$, the average density lies in the range 
$(0.6-4.0) \times 10^9$ cm$^{-3}$ and the corresponding mass range is
$M_{\rm{He}} = 0.01-0.07$ M$_\odot$. The helium line emissivities are taken
from Almog \& Netzer (1989). The density estimates are consistent with 
that estimated by Smith et al. (2008). Based on the observed X-ray luminosities,
Immler et al. (2008) estimate a lower limit to the mass of the X-ray emitting 
shell to be $0.01\rm{M}_\odot$, and the circumstellar density to be 
$\sim 10^7$ cm$^{-3}$. These estimates are much lower than the values implied
by the helium emission lines. This indicates that the X-ray emitting region
could be different from the region emitting the bulk of the helium lines.

\begin{table*}
\caption{He I 5876 and 7065 \AA\ emission line fluxes}
\begin{tabular}{lllllll}
\hline\hline
Phase & \multicolumn{2}{c}{Flux ($10^{-14}$ erg cm$^{-2}$)}& 
\multicolumn{2}{c}{Density$^{\rlap{1}}$ ($10^9$ cm$^{-3}$)}& \multicolumn{2}{c}
{Mass{${^{\rlap{1}}}$} (M$_\odot$)}\\
       & 5876 \AA & 7065 \AA & A$^2$ & B$^3$ & A$^2$ & B$^3$\\
\hline
7.47 & 4.96$^{\rlap{4}}$ & 9.44 & 16.49 & 1.91 & 0.003 & 0.031\\
9.47 & 7.64$^{\rlap{4}}$ & 12.20 & 19.46 & 2.25 & 0.004 & 0.036\\
10.46 & 9.80$^{\rlap{4}}$ & 11.59 & 20.48 & 2.36 & 0.004 & 0.038\\
13.48 & 12.28 & 11.12 & 36.27 & 4.18 & 0.008 & 0.066\\
16.45 & 20.24 & 11.68 & 36.86 & 4.24 & 0.008 & 0.070\\
17.44 & 17.52 & 11.35 & 35.24 & 4.07 & 0.008 & 0.066\\
22.46 & 15.79 & 8.70 & 32.17 & 3.71 & 0.007 & 0.059\\
23.49 & 14.48 & 9.51 & 32.23 & 3.71 & 0.007 & 0.059\\
25.49 & 13.65 & 8.48 & 30.83 & 3.55 & 0.007 & 0.057\\
29.37 & 12.27 & 9.30 & 30.89 & 3.55 & 0.007 & 0.057\\
31.42 & 11.29 & 8.08 & 28.97 & 3.34 & 0.007 & 0.055\\
39.51 & 8.59 & 6.82 & 26.14 & 3.01 & 0.005 & 0.049\\
42.42 & 8.49 & 6.25 & 25.43 & 2.94 & 0.005 & 0.048\\
47.35 & 6.95 & 5.45 & 23.43 & 2.70 & 0.004 & 0.043\\
68.36 & 0.758 & 0.725 & 5.41 & 0.63 & 0.001 & 0.011\\
\hline
\multicolumn{7}{l}{1: Average of 5876 and 7065}\\
\multicolumn{7}{l}{2: Assuming a shell velocity of 600 km s$^{-1}$}\\
\multicolumn{7}{l}{3: Assuming a shell velocity of 2500 km s$^{-1}$}\\
\multicolumn{7}{l}{4: Flux of de-blended narrow component}\\
\end{tabular}
\label{heflx}
\end{table*}

\section{Discussion}

SN 2006jc shows a very peculiar spectrum with a very steep blue continuum and 
dominated by moderately narrow He I emission lines. The presence of the 
moderately narrow emission lines is very similar to that observed in type IIn 
supernovae, where the fast moving supernova ejecta interacts with a 
pre-exisiting circumstellar material. The observed properties of SN 2006jc are 
very similar to the recent supernovae SN 1999cq, SN 2000er and SN 2002ao
(e.g.\ Pastorello et al. 2008). A weak X-ray emission and UV excess have also been 
detected in SN 2006jc, providing further evidence for an interaction with a 
pre-supernova circumstellar material (Immler et al. 2008). A 
luminous mass loss episode was observed in the progenitor of SN 2006jc two 
years prior to the outburst. It is suggested that the strong, intermediate 
width He I emission lines dominating the optical spectrum arise in the
circumstellar shell due to the recent mass loss episode and that it is helium
enriched (Foley et al. 2007, Pastorello et al. 2007, Smith et al. 2008). The 
fluxes of the He I emission lines indicate
a density of $\sim 10^9$~cm$^{-3}$ and a helium mass $\la 0.07$~M$_\odot$ in
the circumstellar shell that is assumed to have a velocity of 2500 km s$^{-1}$
corresponding to the FWHM width of the He I lines. It is also quite likely that 
the velocity of the CSM shell was initially low, and accelerated to 
2500 km s$^{-1}$ due to the interaction. In such a case, the initial velocity 
of the shell is more likely to be in between the assumed velocities, and the
density of the shell $\sim 10^9 - 10^{10}$ cm$^{-3}$. The density estimated 
here is similar to that estimated by Smith et al. (2008). Comparing with 
type IIn SNe, it is found that the estimated density range for SN 2006jc is 
somewhat higher than that estimated for IIn SNe, in which the CSM densities 
are found to range from $\sim 10^6$ (e.g.\ SN 1995N: Fransson et al. 2002) to 
$\ga 10^8$ cm$^{-3}$ (e.g.\ SN 1995G: Pastorello et al. 2002; SN 1997eg: 
Salamanca, Terlevich \& Tenorio-Tagle 2002). 

Dust formation has been observed in SN 2006jc early on, at $\sim 50$ days past 
maximum (Di Carlo et al. 2008, Smith et al. 2008, Nozawa et al. 2008). Dust 
formation is reflected in the helium emission line profiles, which developed 
an asymmetric profile, with the red wing of the profile getting increasingly 
suppressed with time, and also in the increase in the red to NIR continuum
between 65--120 days. Smith et al. (2008) estimate the dust temperature during
this phase to be $\sim 1600$~K. The estimated densities in the shell are high 
enough to precipitate graphite dust (Clayton 1979).

Based on NIR and MIR
observations at $t \sim 200$ days, Sakon et al. (2008) conclude that IR
emission orginated from amorphous carbon grains with two temperatures of
800 K and 320 K. Sakon et al., Nozawa et al. (2008) and Tominaga et al. (2008)
suggest the hot carbon dust is newly formed in the supernova ejecta and heated
by the $^{56}$Ni--$^{56}$Co decay, while the origin of the warm carbon dust is 
a supernova light echo of the CSM carbon dust. For the dust to originate in the
SN ejecta, it implies an ejecta radius of $\sim 10^{16}$~cm during dust
formation. This in turn implies a (constant) velocity $\sim 25,000 - 30,000$~km s$^{-1}$.
No evidence for such a high velocity is seen in the observed spectra. Based on
a comparison of the early spectra of SN 2006jc with those of SN 2000er,
Pastorello et al. (2008) suggest that SN 2006jc was discovered a couple of weeks
after explosion or $\sim 10$ days after maximum light. Very high initial
velocities can thus not be ruled out, as the initial interaction of the
supernova material with the CS material can lead to a deceleration of the 
supernova shell (e.g.\ Chevalier 1982).

On the basis of spectroscopic evidences, Smith et al. propose the 
site of dust formation to be a cold dense shell (CDS) behind the blast wave
and that the shell was composed of dense CSM ejected by the luminous event,
which was then swept up by the forward shock. They, however, do not rule out
the possibility of dust formation in a carbon-rich SN ejecta. Mattila et al.
(2008) also propose the CDS as the site for formation of the hot dust. However, 
they argue, based on the intensity, spectral energy distribution and evolution 
of the IR flux, that the IR emission in SN 2006jc is due to IR echoes. The bulk 
of the near-IR emission is due to an IR echo from the newly formed dust in the 
CDS, while a substantial fraction of the MIR flux is due to pre-existing dust 
in the progenitor wind due to an episodic mass-loss phase that ceased at least 
$\sim 200$ years before the recent pre-supernova luminous outburst and the 
SN event.  

The observed optical light curves of SN 2006jc show an early evolution that is 
quite similar to normal Ib/c supernovae. The initial decline is steep and at 
about 20 days past maximum, the decline slows, a probable indication of the 
supernova having reached the exponential tail. Comparing the light curve 
evolution with type IIn supernovae (e.g.\ SN 1998S), it is seen that the 
early decline is much slower in the case of SNe IIn, where the light curve is 
thought to be powered by the interaction of the supernova material with the CSM 
(e.g.\ Rigon et al. 2003). On the other hand, the light curve evolution of 
SN 2006jc, until the onset of dust formation at $\sim 50$ days since maximum 
is very similar to normal Ib/c objects. Comparing the bolometric light curve of 
SN 2006jc with normal Ib/c SNe and the type IIn SNe, it is seen that the
$uvoir$ bolometric light curve of SN 2006jc is very similar to the normal Ib/c 
objects, with a fast early decline followed by a flattening in the light curve
$\sim 35$ days after maximum.

It is interesting to note that while the spectrum shows clear signatures of
circumstellar interaction similar to type IIn SNe, the bolometric light curve
does not show any evidence of being powered by the interaction. We suggest 
this is possibly a result of a weaker interaction 
in the case of SN 2006jc due to a shell mass that is lower compared to the mass
of the circumstellar material in the case of IIn SNe (eg.\ 
$\sim 0.4$~M$_\odot$ in 1994W: Chugai et al. 2004; $\sim 10$M$_\odot$ in 1997eg:
Salamanca, Terlevich \& Tenorio-Tagle 2002).

\section{Summary}

The optical spectrum and $UBVRI$ light curves of SN 2006jc during the early
phases, until the onset of the dust formation are presented here. The optical
spectrum shows a blue continuum and is dominated by moderately narrow He I
emission lines, similar to type IIn SNe and an indication of the supernova
ejecta interacting with a pre-supernova circumstellar material. The 
moderately narrow He I emission lines arise in the pre-supernova circumstellar 
shell that is helium enriched.

The optical light curves show a clear signature of dust formation as indicated
by a sharp decrease in the magnitudes around day 50, accompanied by a reddening
of the colours. The evolution of the optical light curve during the early phases
is very similar to normal Ib/c SNe. The $uvoir$ bolometric light curve 
evolution of SN 2006jc is reasonably similar to normal Ib/c SNe at all phases.

The He I emission line fluxes indicate the circumstellar shell is dense, with
a density of $\sim 10^9 - 10^{10} $~cm$^{-3}$. The helium mass in this shell
is estimated to be $\la 0.07$~M$_\odot$.

\section*{Acknowledgements}
We thank all the observers of the
2-m HCT (operated by the Indian Institute of Astrophysics), who kindly
provided part of their observing time for the supernova observations.
We thank the referee Andrea Pastorello for indepth comments on the manuscript.
This work has been carried out under the INSA (Indian National Science Academy) -
JSPS (Japan Society for Promotion of Science) exchange programme, and has
also been supported in part by World Premier International
Research Center Initiative (WPI Initiative), MEXT, Japan, and by the
Grant-in-Aid for Scientific Research of the JSPS (18104003, 18540231,
20244035, 20540226) and MEXT (19047004, 20040004).


\begin{thebibliography}{}
\bibitem{} Almog, Y., Netzer, H., 1989, MNRAS, 238, 57
\bibitem{} Arkharov, A., Effinova, N., Leoni, R., Di Paola, A., Di Carlo, E.,
Dolci, M. 2006, ATEL, 961, 1
\bibitem{} Benetti, S., et al. 2006, CBET, 674, 2
\bibitem{} Branch, D. et al. 2002, ApJ, 566, 1005
\bibitem{} Brown, P., Immler, S., Modjaz, M. 2006, ATEL, 916, 1
\bibitem{} Chevalier, R. A. 1982, ApJ, 259, 302
\bibitem{} Chugai, N.N., et al. 2004, MNRAS, 352, 1213
\bibitem{} Chugai, N., Danziger, I.J. 1994, MNRAS, 268, 173
\bibitem{} Clayton, D. D., 1979, Ap\&SS, 65, 179
\bibitem{} Clocchiatti, A. et al. 2001, ApJ, 553, 886
\bibitem{} Crotts, A., Eastman, J., Depoy, D., Prieto, J.L., Garnavich, P.
2006, CBET 672
\bibitem{} Di Carlo, E. et al. 2008, ApJ, 684, 471
\bibitem{} Elmhamdi, A., Danziger, I. J., Cappellaro, E., Della Valle, M., 
Gouiffes, C., Phillips, M. M., Turatto, M.  2004, A\&A, 426, 963
\bibitem{} Fassia, A., et al. 2000, MNRAS, 318, 1093
\bibitem{} Fesen, R., Milisavljevic, D., Rudie, G. 2006, CBET 672, 2
\bibitem{} Fransson, C. et al. 2002, ApJ, 572, 350
\bibitem{} Filippenko, A.V. 1997, ARA\&A, 35, 309
\bibitem{} Filippenko, A.V., Chornock, R., 2002, IAU Circ., 7825, 1
\bibitem{} Foley, R.J., Smith, N., Ganeshalingam, M., Li, W., Chornock, R.,
Filippenko, A.V. 2007, ApJ, L105
\bibitem{} Gehrz, R.D. 1988, ARA\&A, 26, 377
\bibitem{} Immler, S., et al. 2008, ApJ, 674, L85
\bibitem{} Landolt, A.U.,  1992, AJ, 104, 340
\bibitem{} Martin, P., Li, W.D., Qiu, Y.L., West, D. 2002, IAU Circ., 7809, 3
\bibitem{} Matheson, T., Filippenko, A.V., Chornock, R., Leonard, D.C., Li, W.
2000, AJ, 119, 2303
\bibitem{} Matheson, T., Filippenko, A.V., Li, W., Leonard, D.C., Shields, J.C.
2001, AJ, 121, 1648
\bibitem{} Mattila, S., et al. 2008, MNRAS, 389, 141
\bibitem{} Minezaki, T., Yoshii, Y., Nomoto, K. 2007, IAU Circ., 8833, 3
\bibitem{} Nakano, S., Itagaki, K., Puckett, T., Gorelli, R. 2006, CBET, 666, 1
\bibitem{} Nozawa, T. et al. 2008, ApJ, 684, 1343
\bibitem{} Pastorello, A., et al. 2002, MNRAS, 333, 27
\bibitem{} Pastorello, A., et al. 2007, Nature, 447, 829
\bibitem{} Pastorello, A., et al. 2008, MNRAS, 389, 131
\bibitem{} Rigon, L. et al. 2003, MNRAS, 340, 191
\bibitem{} Richmond et al. 1996, AJ, 111, 327
\bibitem{} Sakon, I. et al. 2008, ApJ, in press (arXiv:0711.4801)
\bibitem{} Salamanca, I., Terlevich, R.J., Tenorio-Tagle, G. 2002, MNRAS, 330, 
844
\bibitem{} Schlegel, D.J. 1990, MNRAS, 244, 269
\bibitem{} Smith, N., Foley, R.J., Filippenko, A.V. 2008, ApJ, 680, 568
\bibitem{} Soderberg, A. 2006, ATel, 917, 1
\bibitem{} Stritzinger, M., et al. 2002, AJ, 124, 2100
\bibitem{} Tominaga, N. et al. 2008, ApJ, in press (arXiv:0711.4782)
\end{thebibliography}
\end{document}